\documentclass[A4]{amsart}
\usepackage{amssymb,amsfonts,amsmath,amsthm}

\usepackage{graphicx}

\newcommand{\vp}{\varphi}

\newcommand{\ve}{\varepsilon}

\begin{document}

\title[Boundary value problem]{On the solutions of some boundary value problems for
the general KdV equation}

\author{M. Yu. Ignatyev}
\address{Department of Mathematics,
Saratov University, Saratov, Russia
Astrakhanskaya 83, Saratov 410012, Russia}
\email{mikkieram@gmail.com, ignatievmu@info.sgu.ru}

\begin{abstract}
This paper is concerned with a class of partial differential equations, which are the linear combinations, with constant coefficients, of the classical flows of the KdV hierarchy. A boundary value problem with inhomogeneous boundary conditions of a certain special form is studied. We construct some class of solutions of the problem using the inverse spectral method.
\end{abstract}

\date{}
\maketitle



\section{Introduction}

Boundary and initial-boundary value problems (BVPs and IBVPs) for integrable
nonlinear partial-differential equations play a significant role in mathematical physics being a natural model for the wave processes in semi-bounded space developing under the influence of the boundary regime. The first studies in this area appeared as long ago as 1970-ies, just after the inverse scattering transform (IST) was developed for the Cauchy problems on the
whole line (see, for instance, {\cite{Mos}}). But BVPs were found to be much
more complicated and classical IST failed being applied to them in a straightforward manner.

The further systematic studies on the adaption of the inverse spectral method to the BVPs and IBVPs yield some particular classes of boundary conditions, under which such problems demonstrate some features usually associated with the "integrability". First, to the best of our knowledge, the special role of such particular boundary conditions were mentioned in {\cite{Skl}}, where they were treated from hamiltonian point of view as "the boundary conditions, preserving integrability". In subsequent papers (see, for instance, {\cite{AKhSh}}, {\cite{IBi}}, {\cite{BiTar}}) the BVPs with this special type of boundary conditions were shown to admit the wide classes of exact solutions that can be constructed using the appropriate version of the inverse spectral method.

In a framework of recently developed version of the IST for IBVPs (see {\cite{Fok}}, {\cite{FItsSung}}, {\cite{BdMFSh}} and references therein) the IBVPs with boundary conditions of the above mentioned particular type (now they are known as {\it integrable} or {\it linearizable} boundary conditions) also play a special role. Namely, for such problems it was found possible to reduce the method omitting the step dealing with some nontrivial nonlinear problem. For many classical integrable PDEs the corresponding IBVPs were solved completely {\cite{BiTar}}, {\cite{BiTarNLS}}, {\cite{ISh}}, {\cite{BdMSh}}, also the results on the long--time asymptotics of solutions were obtained (see, for instance, {\cite{FokL}}, {\cite{BdMSh}}).

Although the nature of "integrability" of the boundary conditions is usually clear for each particular integrable PDE some technical aspects of the method require each particular integrable BVP (IBVP) to be considered separately. In this paper, we ask: Can the appropriate version of the inverse spectral method work with some {\it class} of integrable PDEs and some class of BVPs? In this study we start with revisiting the approach and the results of the above mentioned early works such as {\cite{IBi}}, {\cite{AKhSh}} and examine the set of PDEs, associated in a framework of IST with a classical Sturm--Liouville spectral problem.
The most known in this family is the classical KdV equation:
$$
q_t=6qq_x-q_{xxx}. \eqno(1.1)
$$
In {\cite{AGKh}} the following boundary conditions:
$$
q(0,t)=a, \ q_{xx}(0,t)=b \eqno(1.2)
$$
with real constants $a,b$ were shown to be the only integrable boundary conditions for (1.1). The corresponding BVP was investigated in several works (see, for instance {\cite{AKhSh}}, {\cite{FokL}}, {\cite{MYI}}). In particular, in {\cite{MYI}} the following result was obtained. Given the function $p(t)$ from a certain class and the number $w_0$ from the certain interval, define $M(T,\lambda)$ as a Weyl--Titchmarsh function for the Sturm--Liouville operator with the potential $p(t+T)$ and calculate $w(t)$ as a solution for the Cauchy problem $\dot w+w^2=p(t)$, $w(0)=w_0$. Then the function $m(t,\lambda)$ defined as:
$$
m(t,\lambda)=\frac{M\left(t,f(\lambda)\right)-w(t)}{4\lambda+2a}, \ f(\lambda)=16\lambda^{3}-(12a^{2}-4b)\lambda-2a(2a^{2}-b) \eqno(1.3)
$$
is a Weyl-Titchmarsh function for some Sturm-Liouville operator $L(t)=-d^2/dx^2+q(x,t)$ and $q(x,t)$ is a solution of the BVP (1.1), (1.2).

In this paper, we consider the BVP for a class of partial differential equations, which are the linear combinations, with constant coefficients, of the classical flows of the KdV hierarchy ({\it general KdV equations}). Under some condition on the mentioned constant coefficients and certain special choice of the boundary conditions we provide the extension of (1.3) that generates some class of solutions for this BVP. The exact formulation of our main result is contained in the Theorem 4.1. Here we notice that the constructed class of solutions for the BVP contains, in particular, soliton and finite-gap solutions. Actually, we work with a wider class $\tilde B$ introduced by V.A. Marchenko (see {\cite{march}} for detailed discussion on the corresponding spectral theory and it's application to the KdV equation).

\section{Some preliminary facts and notations}

Consider on the real axis $-\infty<x<\infty$ the Sturm--Liouville operator of the form:
$$
L=-\frac{d^2}{dx^2}+q(x) \eqno(2.1)
$$
with bounded real--valued potential $q$. Let $\psi^\pm(x,\lambda)$ be the Weyl solutions for $L$ on the semi-axes $(0,\pm\infty)$ normalized as $\psi^\pm(0,\lambda)=1$ and let $m_\pm(\lambda):=(\psi^\pm) '(0,\lambda)$ be the corresponding Weyl-Titchmarsh functions. For infinitely differentiable $q$, the following asymptotic expansion holds:
$$
m_+(\lambda)=i\rho+\sum\limits_{n=1}^\infty \frac{b_n}{(i\rho)^n}, \quad \lambda\to\infty, \arg \lambda\in[\ve,\pi-\ve], \lambda=\rho^2, \mbox{Im}\rho>0. \eqno(2.2)
$$
The coefficients $b_n=b_n(q)$ can be considered as some nonlinear functionals with respect to the potential $q$. It is considerable that $b_n(q)$ are actually polynomials of $q$ and it's derivatives at $x=0$. Indeed, they can be calculated (see, for instance, {\cite{Lev1}}) as $b_n=2^{-n}\beta_n(0)$, where:
$$
\beta_1=q, \quad \beta_{n+1}=-\beta'_n-\sum\limits_{\nu=1}^{n-1}\beta_{\nu}\beta_{n-\nu}.
$$

Now we recall some facts about the rapidly decreasing reflectionless potentials (see, for instance {\cite{march}}).

Denote by
$B(\mu)$, $\mu<0$ the set of all reflectionless potentials $q$ such that the spectrum of the operator $L$
is located on $[\mu,\infty)$ and define $B:=\bigcup\limits_{\mu<0}B(\mu)$.

Let $e^{\pm}(x,\rho), \ \pm\mbox{Im}\rho\geq 0$ be the Jost solutions for $L$ normalized with the asymptotics:
$$
e^{\pm}(x,\rho)=\mbox{e}^{i\rho x}(1+o(1)), \ x\to\pm\infty
$$
and $\psi(x,\rho)$ be the {\it Weyl--Marchenko solution} defined as follows:
$$
\psi(x,\rho):=\psi^\pm(x,\rho^2)=\frac{e^{\pm}(x,\rho)}{e^{\pm}(0,\rho)}, \ \pm\mbox{Im}\rho>0.
$$
The function $m(\rho):=\psi'(0,\rho)=m_\pm(\rho^2)$, $\pm\mbox{Im}\rho>0$ is called the {\it Weyl--Marchenko function}. If $q\in B(-a^2)$, $a>0$ then the Weyl--Marchenko function admits the representation:
$$
m(\rho)=i\rho+i\int\limits_{-a}^a\frac{d\sigma(\xi)}{\rho-i\xi}, \eqno(2.3)
$$
where $d\sigma$ is a discrete measure concentrated at the finite set of points $\Lambda_0(q)$ $=\{\xi_k\}_{k=1}^{n_0}$. Inversely, any function $m(\rho)$ of the form (2.3) is a Weyl--Marchenko function for some $q\in B$; moreover $q\in B(-\mu^2)$, $\mu^2=a^2+\int_{-a}^a d\sigma(\xi)$.

Let $\{-\kappa^2_k\}_{k=1}^n, \ \kappa_k>0$ be the set of eigenvalues of $L$. Define $\Lambda(q)=\{\kappa_k\}_{k=1}^n$. It is known that $n_0=n$ and the ordering can be chosen such that
$$
0\leq\left|\xi_1\right|\leq\kappa_1\leq\left|\xi_2\right|\leq\ldots\leq\left|\xi_n\right|\leq\kappa_n.
$$
Moreover, if we split $\Lambda(q)=\Lambda_1(q)\cup\Lambda_2(q)$, $\Lambda_2(q)=\Lambda(q)\cap\Lambda_0(q)$, $\Lambda_1(q)=\Lambda(q)\setminus\Lambda_2(q)$ then for any $\xi\in\Lambda_2(q)$ we necessarily have $-\xi\in\Lambda_0$ and inversely: if $\xi\neq 0$ and both $\pm\xi\in\Lambda_0$ then $|\xi|\in\Lambda_2$. Also one can notice that the set $\{\rho=\pm i\kappa :\kappa\in \Lambda_1(q)\}$ coincides with the set of nonzero roots of the equation:
$$
m(\rho)=m(-\rho). \eqno(2.4)
$$

For $\kappa\in\Lambda(q)$ we denote as $\alpha(\kappa)$ the normalizing constant:
$$
e^-(x,-i\kappa)=\alpha(\kappa)e^+(x,i\kappa). \eqno(2.5)
$$
Normalizing constants can be represented in terms of $\Lambda(q)$, $\Lambda_0(q)$ and $d\sigma$ as follows:
$$
\alpha(\kappa)=\prod_{\xi\in\Lambda^1_0(q)}\frac{\kappa+\xi}{\kappa-\xi}, \ \kappa\in\Lambda_1(q), \eqno(2.6)
$$
and
$$
\alpha(\kappa)=-\frac{d\sigma(-\kappa)}{d\sigma(\kappa)} \prod_{\xi\in\Lambda^1_0(q)}\frac{\kappa+\xi}{\kappa-\xi}, \ \kappa\in\Lambda_2(q), \eqno(2.7)
$$
where $\Lambda^1_0(q):=\left\{\xi\in\Lambda_0(q): \ |\xi|\notin\Lambda_2(q)\right\}$.

Now we consider the set $\tilde B=\bigcup_{\mu<0}\overline{B(\mu)}$, where the closure is considered in the topology of uniform convergence of the functions on
each compact set of the real axis. For $q\in \tilde B$ we define as above the Weyl--Marchenko solution $\psi(x,\rho)$ as
$$
\psi(x,\rho)=\psi^{\pm}(x,\rho^2), \ \pm\mbox{Im}\rho>0,
$$
and the Weyl--Marchenko function $m(\rho)$ as $m(\rho):=\psi'(0,\rho)$. If $q\in\overline{ B(-a^2)}$, $a>0$ then the Weyl--Marchenko function admits the representation (2.3) with some measure $d\sigma$ concentrated on $[-a,a]$ and satisfying the estimate $\int_{-a}^a d\sigma(\xi)<a^2$. Inversely, any function of the form (2.3) with an arbitrary measure $d\sigma$ is a Weyl--Marchenko function for some $q\in \tilde B$;  moreover $q\in \overline{B(-\mu^2)}$, $\mu^2=a^2+\int_{-a}^a d\sigma(\xi)$. It is clear from the representation (2.3) that for $q\in\tilde B$ the corresponding Weyl--Marchenko function $m(\rho)$ is holomorphic outside some finite segment of the imaginary axis, the corresponding Laurent series has the same form as an asymptotic expansion (2.2), namely:
$$
m(\rho)=i\rho+\sum\limits_{n=1}^\infty \frac{b_n(q)}{(i\rho)^n}. \eqno(2.8)
$$

We complete our discussion of the classes $B$ and $\tilde B$ with mentioning the properties that follows from the arguments used in proof of {\cite{march}}, Theorem 2.1.

\medskip
{\bf Proposition 2.1.} Let $q\in\overline{B(-a^2)}$. Then for any sequence $q_N\in B(-a^2)$ convergent to $q$ in the topology of $\tilde B$ there exists the subsequence $q_{N_n}$ such that the corresponding Weyl--Marchenko functions $m_{N_n}(\rho)$ converge to $m(\rho)$ for all $\rho\in\mathbf{C}\setminus[-ia,ia]$. The Weyl--Marchenko solutions $\psi_{N_n}(x,\rho)$ also converge to $\psi(x,\rho)$ (uniformly in $x$ on each finite segment).

\medskip
{\bf Remark 2.1.} Under the conditions of the Proposition 2.1, the following simple but useful assertion is true.
Take an arbitrary $\rho\in\mathbf{C}\setminus[-ia,ia]$ and let $f_n(x)$ be the solutions of the Cauchy problems:
$$
-f''_n+q_{N_n}(x)f_n=\rho^2 f_n, \ f_n(0)=f^0_n,\ f'_n(0)=f^1_n,
$$
where $\lim_{n\to\infty} f_n^0=f^0$, $\lim_{n\to\infty} f_n^1=f^1$. Then $\lim_{n\to\infty} f_n^{(\nu)}(x)=f^{(\nu)}(x)$, $\nu=0,1$, where $f(x)$ is a solution of the Cauchy problem:
$$
-f''+q(x)f=\rho^2 f, \ f(0)=f^0,\ f'_n(0)=f^1.
$$

\medskip
Now we recall the construction of the classical KdV flows and describe the equations that will be considered in this paper. Let us define the $X_\nu=X_\nu(q)$, which are the polynomials of $q$ and its derivatives constructed via the following recurrent procedure:
$$
\ P_1=-\frac{1}{2}q, \ P'_{\nu+1}=HP_\nu,
$$
$$
H=-\frac{1}{2}\frac{d^3}{dx^3}+2q\frac{d}{dx}+q',
$$
$$
X_\nu:=-P'_{\nu+1}.
$$
The $s$-th KdV flow has the form:
$$
\dot q=X_s(q)
$$
(here and below "dot" denotes the derivative in $t$ while "prime" denotes the derivative with respect to $x$). It is well-known that this equation can be integrated via the classical IST, namely the evolution of the scattering data of the associated Sturm--Liouville operator $L(t)=-d^2/dx^2+q(x,t)$ on the whole line has the form:
$$
r(\rho,t)=r(\rho,0)\mbox{e}^{2i\vp(\rho)t}; \ \alpha(\kappa,t)=\alpha(\kappa,0)\mbox{e}^{2i\vp(i\kappa)t},
$$
where $r(\rho,t)$ is a reflection coefficient, $\alpha(\kappa,t)$ is a normalizing constant corresponding to an (time-independent) eigenvalue $-\kappa^2$ and $\vp(\rho)=2^{s-1}\rho^{2s+1}$. In this paper, we consider the following "general KdV equation" (see, for instance, {\cite{lev2}}):
$$
\dot q=\sum\limits_{\nu=0}^s C_{\nu} X_\nu(q).
$$
This equation can also be integrated using the IST and the evolution of the scattering data of the associated Sturm--Liouville operator has the form {\cite{lev2}}:
$$
r(\rho,t)=r(\rho,0)\mbox{e}^{2i\vp(\rho)t}; \ \alpha(\kappa,t)=\alpha(\kappa,0)\mbox{e}^{2i\vp(i\kappa)t},
$$
where
$$
\varphi(\rho)=\frac{1}{2}\rho\sum\limits_{\nu=0}^s C_\nu \left(2\rho^2\right)^\nu. \eqno(2.9)
$$

\section{Formulation of the Problem}

We consider the nonlinear partial-differential equation of the following form ("general KdV equation", see, for instance, {\cite{lev2}}):
$$
\dot q=\sum\limits_{\nu=0}^s C_{\nu} X_\nu(q), \eqno(3.1)
$$
where $C_\nu$ are real constants together with the boundary conditions:
$$
b_{2n}(q(\cdot,t))=0,\ n=\overline{1,s-1}, \ b_{2n-1}(q(\cdot,t))=a_n,\ n=\overline{1,s+1}. \eqno(3.2)
$$
Here $X_\nu(q)$ and $b_n(q)$ were defined in a previous section while $a_n, n=\overline{1,s+1}$ are the real constants, such that $a=(a_1,\dots,a_{s+1})\in \mathcal{A}$, where $\mathcal{A}$ is a certain one-parametric set in $\mathbf{R}^{s+1}$ that we describe below in this section.

For example, for $s=1$ we consider the equation:
$$
\dot q= \frac{1}{2}C_0 q'+\frac{1}{4}C_1\left(6qq'-q'''\right)
$$
together with the boundary conditions:
$$
\frac{1}{2}q(0,t)=a_1,  \ \frac{1}{8}(q''(0,t)-q^2(0,t))=a_2;
$$
for $s=2$ we deal with the BVP of the following form:
$$
\dot q= \frac{1}{2}C_0 q'+\frac{1}{4}C_1\left(6qq'-q'''\right)+\frac{1}{8}C_2\left(q^{(5)}-10qq'''-20q'q''+30q^2q'\right),
$$
$$
\frac{1}{2}q(0,t)=a_1, \ q'(0,t)=0, \ \frac{1}{8}(q''(0,t)-q^2(0,t))=a_2, $$ $$ \frac{1}{32}\left(q^{(4)}(0,t)-6q(0,t)q''(0,t)-5(q'(0,t))^2+2q^3(0,t)\right)=a_3.
$$

In the sequel we assume that the constants $C_\nu$ satisfy the following constraint.

\medskip
{\bf Assumption 1.} The polynomial $\vp(\rho)$ defined in (2.9)
can be written in the following form:
$$
\varphi(\rho)=4^s\rho\prod_{\nu=1}^s\left(\rho^2-d_\nu\right)=4^s\rho\prod_{\nu=1}^s\left(\rho^2+\delta^2_\nu\right) \eqno(3.3)
$$
with $0<\delta_1<\ldots<\delta_s$.

\medskip
Now let us specify the set $\mathcal{A}$.
Define the polynomial:
$$
f(\lambda):=16^s\lambda\prod_{\nu=1}^s\left(\lambda-d_\nu\right)^2,
$$
i.e., such that $(\varphi(\rho))^2=f(\rho^2)$, and consider the equation:
$$
f(\lambda)=\mu. \eqno(3.4)
$$
Let $\mu^-$ be the greatest lower bound of the set of all real $\mu$ such that all the roots of (3.4) are real. Take an arbitrary $\mu\in(\mu^-,0)$. Consider (3.4) and denote it's roots as $0>c_0>c_1>c'_1>\ldots>c_s>c'_s$ (see figure 1). (In the sequel it will be convenient to assume that $c_\nu$ were defined as the roots of $f(\lambda)$ satisfying the condition: $f'(c_\nu)<0$).
Define the polynomial
$$
g_\mu(\lambda):=4^s\prod_{\nu=1}^s\left(\lambda-c_\nu\right) \eqno(3.5)
$$
and the real numbers $a_n(\mu)$ as coefficients of the following Laurent series:
$$
\frac{4^s\prod\limits_{\nu=1}^s(\lambda-d_\nu)}{g_{\mu}\left(\lambda\right)}=\prod\limits_{\nu=1}^s\frac{\lambda-d_\nu}{\lambda-c_\nu}=
1+\sum\limits_{n=1}^\infty\frac{(-1)^n a_n(\mu)}{\lambda^n}. \eqno(3.6)
$$
We set $a(\mu)=(a_1(\mu),\dots,a_{s+1}(\mu))$ and $\mathcal{A}=\{a(\mu), \mu\in(\mu^-,0)\}$.

\begin{figure}[h]
\center{\includegraphics{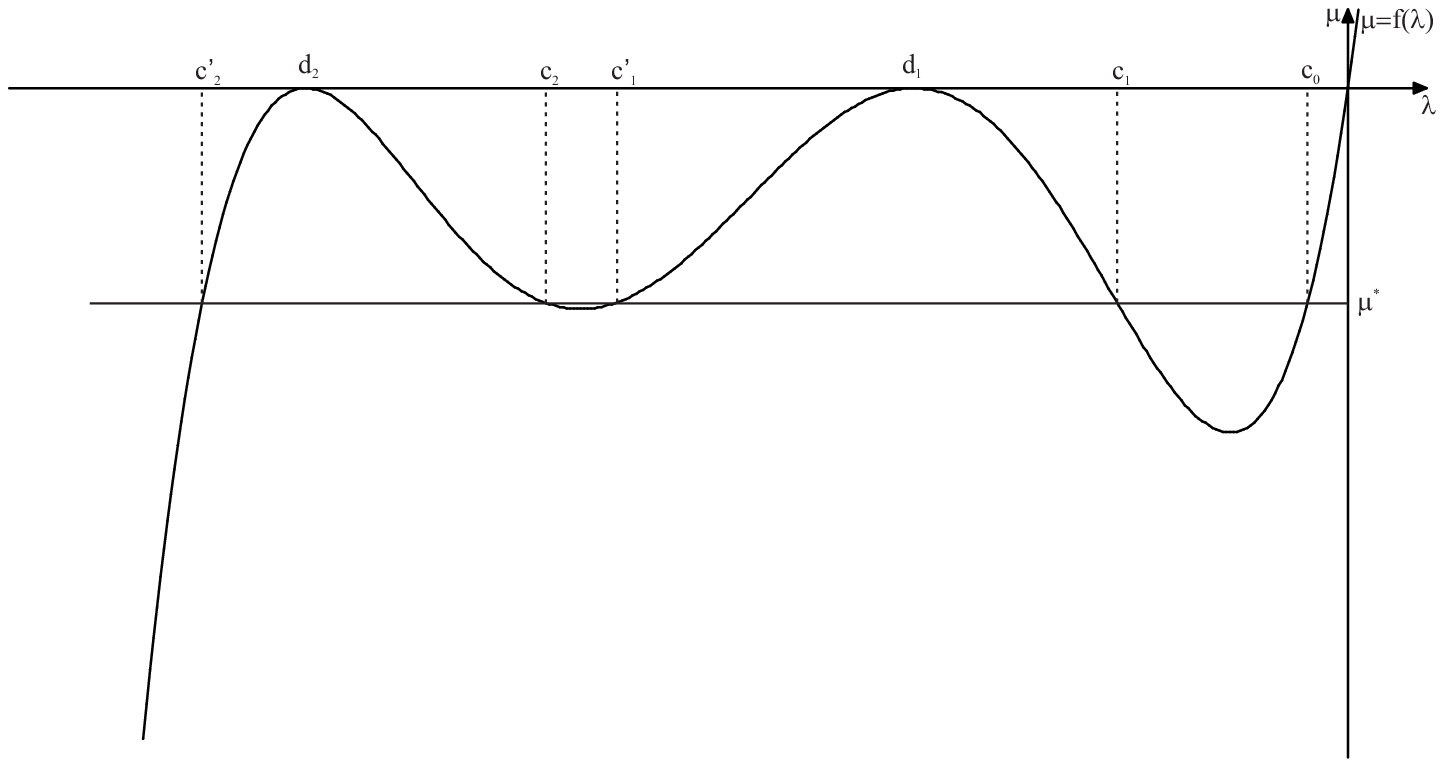}}
\caption{}
\label{ris:figure1}
\end{figure}

\section{Main result}

Let us choose an arbitrary $\mu^*\in(\mu^-,0)$ and consider the problem (3.1), (3.2) with $(a_1,\dots,a_{s+1})=a(\mu^*)\in \mathcal{A}$.

\medskip
{\bf Theorem 4.1.} {\it
 Let $Q$ be an arbitrary function from $\overline{B(\mu_*)}$, $\mu_*\in(\mu^*,0)$. Denote by $M(T,\cdot)$,  the Weyl--Marchenko function for $Q_T(t):=Q(t+T)$. Let $w$ be a solution of the Cauchy problem:
$$
\dot{w}+w^2=Q(t)-\mu^*, \quad w(0)=w_0 \eqno(4.1)
$$
with an arbitrary $w_0\in(M(0, i\kappa^*),M(0, -i\kappa^*))$, $\kappa^*:=\sqrt{-\mu^*}$. Denote $g(\lambda)=g_{\mu^*}(\lambda)$ (where  $g_{\mu}(\lambda)$ is the polynomial defined via (3.5)).

Then the function $m(t,\cdot)$ defined as:
$$
m(t,\rho):=\frac{M(t,\varphi(\rho))-w(t)}{g\left(\rho^2\right)}, \eqno(4.2)
$$
is a Weyl--Marchenko function for some function $q(\cdot,t)\in \tilde B$ and the function $q(x,t)$ is a solution of the boundary value problem (3.1), (3.2) on each of the two semi-axes $(-\infty,0)$, $(0,\infty)$.
}

\medskip
{\bf Proof.} We start with the following remark. Since $-{\kappa^*}^2<-\kappa_*^2$ the value $-{\kappa^*}^2$ cannot be a Dirichlet eigenvalue for the Sturm-Liouville operator with the potential $Q$ on any of semi-axes $(0,\pm\infty)$. Therefore both $M(0, i\kappa^*),M(0, -i\kappa^*)$ exist and are  finite. Moreover, one can easily show that $M(0,i\kappa^*)<M(0,-i\kappa^*)$ (it is clear for $q\in B(\mu_*)$, for $q\in \overline{B(\mu_*)}$ it can be proved via the limiting procedure). This means that the interval $(M(0, i\kappa^*),M(0, -i\kappa^*))$ specified in the Theorem is not empty.
Similarly for any $T$ $-{\kappa^*}^2$ is not a Dirichlet eigenvalue for the Sturm-Liouville operator with the potential $Q_T$ therefore $\psi(T,\pm i\kappa^*)\neq 0$ and $M(T,\pm i\kappa^*)=\dot\psi(T,\pm i\kappa^*)/\psi(T,\pm i\kappa^*)$ are finite.
The comparison theorem for the Riccati equation yields the estimate $M(T,i\kappa^*)<w(T)<M(T,-i\kappa^*)$, and since  $M(T,\pm i\kappa^*)$ are finite for all $T$ we can conclude the $w(T)$ is also finite for all $T$. This means, in particular, that the function $m(t,\rho)$ is correctly defined via (4.2).

The further proof will be divided into several steps.

\medskip
{\bf Lemma 4.1.} Under the conditions of Theorem for any fixed $t$ $m(t,\cdot)$ is a Weyl--Marchenko function for some $q(\cdot,t)\in\overline{B(-\tau_*^2)}$, where $\tau_*$ depends only upon $\mu_*$ and $\mu^*$. Moreover, $q(\cdot,t)$ satisfies the boundary conditions (3.2).

\medskip
{\bf Proof of Lemma 4.1.}
Our plan is to obtain the representation of the form (2.3) for $m(t,\rho)$. Throughout this calculations the parameter $t$ is arbitrary but fixed and for the sake of brevity we omit it in all the arguments.

First we use the relation
$$
\frac{1}{g\left(\rho^2\right)}=\sum\limits_{\nu=1}^{s}\frac{1}{g'(c_\nu)(\rho^2-c_\nu)}=\sum\limits_{\nu=1}^{s}\frac{i}{2\gamma_\nu g'(c_\nu)}\left[-\frac{1}{\rho-i\gamma_\nu}+\frac{1}{\rho+i\gamma_\nu}\right],
$$
where $\gamma_\nu=\sqrt{-c_\nu}$ and rewrite (4.2) into the following form:
$$
m(\rho)=\sum\limits_{\nu=1}^s\frac{i}{2\gamma_\nu g'(c_\nu)}\left[\frac{w-M(\vp(\rho))}{\rho-i\gamma_\nu}+\frac{M(\vp(\rho))-w}{\rho+i\gamma_\nu}\right]
$$
Then we note that $\vp(\pm i\gamma_\nu)=\pm (-1)^{\nu-1}i\kappa^*$ and rewrite this representation as follows:
$$
m(\rho)=m_1(\rho)+m_2(\rho), \eqno(4.3)
$$
where
$$
m_1(\rho)=\sum\limits_{\nu=1}^s\frac{i}{2\gamma_\nu g'(c_\nu)}\left[\frac{M\left(\vp(i\gamma_\nu)\right)-M(\vp(\rho))}{\rho-i\gamma_\nu}+
\frac{M(\vp(\rho))-M\left(\vp(-i\gamma_\nu)\right)}{\rho+i\gamma_\nu}\right], \eqno(4.4)
$$
$$
m_2(\rho)=
\sum\limits_{\nu=1}^s\frac{i}{2\gamma_\nu g'(c_\nu)}\left[\frac{w-M((-1)^{\nu-1}i\kappa^*)}{\rho-i\gamma_\nu}+\frac{M((-1)^{\nu}i\kappa^*)-w}{\rho+i\gamma_\nu}\right]. \eqno(4.5)
$$

Now we use the representation (2.3) for $M(t,\cdot)$:
$$
M(\mu)=i\mu+i\int\limits_{-\kappa_*}^{\kappa_*}\frac{d\theta(\eta)}{\mu-i\eta},
$$
where $\kappa_*=\sqrt{-\mu_*}$ to rewrite (4.4) into the following form:
$$
m_1(\rho)=m_{10}(\rho)+\sum\limits_{\nu=1}^s\frac{i}{2\gamma_\nu g'(c_\nu)}\left\{\frac{i}{\rho-i\gamma_\nu}\int\limits_{-\kappa_*}^{\kappa_*} \left[ \frac{1}{\vp(i\gamma_\nu)-i\eta}-\frac{1}{\vp(\rho)-i\eta}\right]d\theta(\eta)+\right. $$ $$ \left.\frac{i}{\rho+i\gamma_\nu}\int\limits_{-\kappa_*}^{\kappa_*} \left[ \frac{1}{\vp(\rho)-i\eta}-\frac{1}{\vp(-i\gamma_\nu)-i\eta}\right]d\theta(\eta)\right\}, \eqno(4.6)
$$
where
$$
m_{10}(\rho)=\sum\limits_{\nu=1}^s\frac{i}{2\gamma_\nu g'(c_\nu)}\left\{i\frac{\vp(i\gamma_\nu)-\vp(\rho)}{\rho-i\gamma_\nu}+ i\frac{\vp(\rho)-\vp(-i\gamma_\nu)}{\rho+i\gamma_\nu}\right\}. \eqno(4.7)
$$
Now consider (4.7) in details. One can easily notice that its right-hand side is a polynomial while taking the limit as $\rho\to\infty$ we obtain:
$$
m_{10}(\rho)=\sum\limits_{\nu=1}^s\frac{i}{2\gamma_\nu g'(c_\nu)}\left[\frac{i\vp(\rho)}{\rho+i\gamma_\nu}- \frac{i\vp(\rho)}{\rho-i\gamma_\nu}\right]+O\left(\frac{1}{\rho}\right)=i\frac{\vp(\rho)}{g\left(\rho^2\right)}+O\left(\frac{1}{\rho}\right)= i\rho+O\left(\frac{1}{\rho}\right).
$$
Thus we have $m_{10}(\rho)=i\rho$ and we can return to (4.6) and write it in the following way:
$$
m_1(\rho)=i\rho+\int\limits_{-\kappa_*}^{\kappa_*}\sum\limits_{\nu=1}^s\frac{i}{2\gamma_\nu g'(c_\nu)}\left\{\frac{i}{\rho-i\gamma_\nu} \cdot \frac{\vp(\rho)-\vp(i\gamma_\nu)}{(\vp(i\gamma_\nu)-i\eta)(\vp(\rho)-i\eta)}\right. $$ $$ \left. +\frac{i}{\rho+i\gamma_\nu} \cdot \frac{\vp(-i\gamma_\nu)-\vp(\rho)}{(\vp(-i\gamma_\nu)-i\eta)(\vp(\rho)-i\eta)}\right\}d\theta(\eta).
$$
Note that the integrand is a sum of some  meromorphic functions vanishing at infinity with poles that are the roots of the equation $\vp(\rho)=i\eta$. Clear that for $\eta\in[-\kappa_*,\kappa_*]$ all these roots are pure imaginary and can be written in the form $i\xi_j=i\xi_j(\eta), \ j=\overline{-s,s}$, where $\xi_0\in(-\gamma_0,\gamma_0)$, $\xi_j\in(\gamma_j, \gamma'_j),\ j=\overline{1,s}$, $\xi_{-j}\in(-\gamma'_j, -\gamma_j),\ j=\overline{1,s}$ (see figure 2, we recall that $\gamma_j=\sqrt{-c_j}$ and define $\gamma'_j=\sqrt{-c'_j}$).

\begin{figure}[h]
\center{\includegraphics{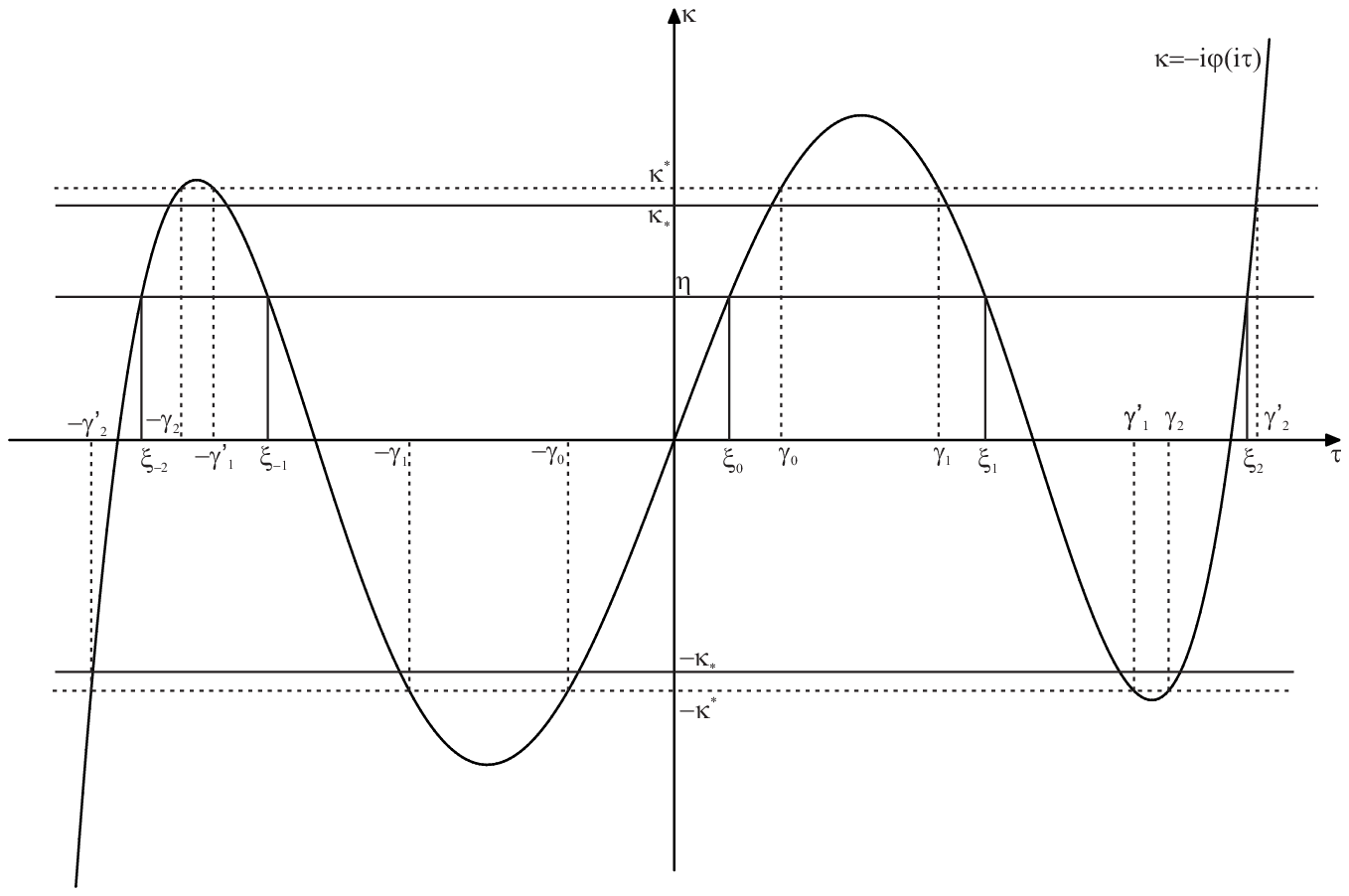}}
\caption{}
\label{ris:figure2}
\end{figure}

Using the representations:
$$
\frac{i}{\rho-i\gamma_\nu} \cdot \frac{\vp(\rho)-\vp(i\gamma_\nu)}{(\vp(i\gamma_\nu)-i\eta)(\vp(\rho)-i\eta)}= \sum\limits_{j=-s}^s \frac{1}{\vp'(i\xi_j)(\rho-i\xi_j)}\cdot\frac{1}{\gamma_\nu-\xi_j},
$$
$$
\frac{i}{\rho+i\gamma_\nu} \cdot \frac{\vp(-i\gamma_\nu)-\vp(\rho)}{(\vp(-i\gamma_\nu)-i\eta)(\vp(\rho)-i\eta)}= \sum\limits_{j=-s}^s \frac{1}{\vp'(i\xi_j)(\rho-i\xi_j)}\cdot\frac{1}{\gamma_\nu+\xi_j}
$$
we obtain:
$$
m_1(\rho)=i\rho+i \int\limits_{-\kappa_*}^{\kappa_*}\sum\limits_{\j=-s}^s\frac{1}{\vp'\left(i\xi_j(\eta)\right)(\rho-i\xi_j(\eta))} \sum\limits_{\nu=1}^s\frac{1}{2\gamma_\nu g'(c_\nu)}\left\{\frac{1}{\gamma_\nu-\xi_j(\eta)}+\frac{1}{\gamma_\nu+\xi_j(\eta)}\right\}d\theta(\eta)
$$
that yields after some algebra:
$$
m_1(\rho)=i\rho+i\sum\limits_{j=-s}^{s}\int\limits_{-\kappa_*}^{\kappa_*} \frac{d\theta(\eta)}{\left(\rho-i\xi_j(\eta)\right)\vp'\left(i\xi_j(\eta)\right)g\left(-\xi^2_j(\eta)\right)}. \eqno(4.8)
$$
In each integral we make a change of variable $\xi_j(\eta)=\xi$ and arrive at:
$$
m_1(\rho)= i\rho+i\sum\limits_{j=-s}^{s}\int\limits_{I_j} \frac{d\sigma_j(\xi)}{\rho-i\xi}, \eqno(4.9)
$$
where $I_j, \ j=\overline{-s,s}$ are the segments $I_j\subset (\gamma_j,\gamma'_j), \ I_{-j}\subset (-\gamma'_j,-\gamma_j)$, $j=\overline{1,s}$, $I_0\subset(-\gamma_0,\gamma_0)$ and $\sigma_j(\xi)$ are nondecreasing functions on $I_j$ defined as:
$$
d\sigma_j(\xi)=\frac{(-1)^j d\left(\theta(-i\vp(i\xi))\right)}{\vp'(i\xi)g(-\xi^2)}. \eqno(4.10)
$$
Now let us return to the representation (4.5). From the estimate $M(i\kappa^*)<w<M(-i\kappa^*)$ mentioned above we get $\mbox{sgn}\left(w-M((-1)^{\nu-1}i\kappa^*)\right)$ $=\mbox{sgn}\left(M((-1)^{\nu}i\kappa^*)-w\right)$ $=\mbox{sgn}g'(c_\nu)=(-1)^{\nu-1}$. This means that in the representation:
$$
m_2(\rho)= i\int\limits_{-\gamma'_s}^{\gamma'_s}\frac{d\sigma^0(\xi)}{\rho-i\xi}, \eqno(4.11)
$$
$\sigma^0$ corresponds to a discrete measure concentrated at the points $\pm\gamma_\nu, \ \nu=\overline{1,s}$ and
$$
d\sigma^0(\pm\gamma_\nu)=\frac{\pm\left(w-M(\pm(-1)^{\nu-1}i\kappa^*)\right)}{2\gamma_\nu g'(c_\nu)}>0. \eqno(4.12)
$$
Finally, gathering together (4.3), (4.9)-(4.12) we can conclude that $m(\rho)$ can be represented in the form:
$$
m(\rho)=i\rho+ i\int\limits_{-\gamma'_s}^{\gamma'_s}\frac{d\sigma(\xi)}{\rho-i\xi}, \eqno(4.13)
$$
where
$$
\sigma(\xi)=\sigma^0(\xi)+\sum\limits_{j=-s}^s\chi_j(\xi)\sigma_j(\xi), \eqno(4.14)
$$
and $\chi_j$ are the characteristic functions of the segments $I_j$.

From the representation (4.13) and the arguments above it follows that $m(t,\cdot)$ is a Weyl--Marchenko function for some $q(\cdot,t)\in \overline {B(-\tau_*^2)}$ with some $\tau_*$. Let us show that $\tau_*$ can be chosen independent on $t$ and moreover independent on the particular choice of $Q\in\overline{B(\mu_*)}$.
For this purpose we evaluate the measure $d\sigma$ defined in (4.10), (4.12).

First we note that the measures $\chi_j d\sigma_j$ are concentrated on the segments $I_j$ that do not contain the zeros of denominators in (4.10). Moreover, the endpoints of the segments $I_j$ are all among the points of the set $\{\xi_j(\pm\kappa_*)\}_{j=\overline{-s,s}}$ and since $\kappa_*<\kappa^*$ they are on some positive distance depending only upon $\mu^*$ and $\mu_*$ from $\gamma_j$, $\gamma'_j$ and the extremal points of $\vp$, i.e. from the zeros of the above mentioned denominators. This means that all the denominators in (4.10) can be estimated from below by some positive constant that depends on $\mu_*$, $\mu^*$ but not on the function $Q$ and the parameter $t$. Thus we can estimate
$$
\int\limits_{I_j}d\sigma_j(\xi)<C\int\limits_{-\kappa_*}^{\kappa_*} d\theta(\eta). \eqno(4.15)
$$
Further, using, for instance, {\cite{march}}, Lemma 2.2 one can show that for any $Q\in\overline{B(\mu_*)}$ the Weyl--Marchenko function $M(t,\mu)$ is bounded for $t\in(-\infty,\infty)$ and any fixed $|\mu|>\sqrt{|\mu_*|}$ with some constant that depends only upon $\mu$, $\mu_*$ and not on $Q\in\overline{B(\mu_*)}$. This means that $|w(t)-M(t,\pm i\kappa^*)|<C$, i.e.
$$
\int\limits_{-\gamma'_s}^{\gamma'_s}d\sigma^0(\xi)<C \eqno(4.16)
$$
with some constant $C$ that does not depend on $Q$ and $t$.

Since $\int_{-\kappa_*}^{\kappa_*} d\theta(\eta)$ itself is bounded with some constant depending only upon $\mu_*$ (see Section 2) we get the estimate $$
\int\limits_{-\gamma'_s}^{\gamma'_s}d\sigma(\xi)<C=C(\mu^*,\mu_*)
$$
that yields that $\tau_*$ actually depends only upon $\mu^*$, $\mu_*$.

Finally, we compare the Laurent series (2.8) for $m(t,\cdot)$ with the expansion obtained from (4.2) and the asymptotics $M(t,\rho)=i\rho+O\left(\rho^{-1}\right)$. Namely, from (4.2) we obtain:
$$
m(t,\rho)=i\frac{\vp(\rho)}{g\left(\rho^2\right)}-\frac{w(t)}{4^s\rho^{2s}}+O\left(\rho^{-2s-2}\right).
$$
Taking into account (3.3), (3.5) and the expansion (3.6) we rewrite this in the form:
$$
m(t,\rho)=i\rho\left(1+\sum\limits_{n=1}^{s+1}\frac{a_n}{(i\rho)^{2n}}\right)-\frac{w(t)}{4^s\rho^{2s}}+O\left(\rho^{-2s-2}\right),
$$
while (2.8) reads as
$$
m(t,\rho)=i\rho+\sum\limits_{n=1}^\infty \frac{b_n(q(\cdot,t))}{(i\rho)^n}
$$
and we arrive at (3.2).
$\hfil\Box$

\medskip
{\bf Lemma 4.2.} Let $Q$ in the conditions of Theorem 4.1 be a reflectionless potential from $B(\mu_*)$. Then corresponding $q(\cdot,\cdot)$ satisfies the equation (3.1).

\medskip
{\bf Proof of Lemma 4.2.}
Consider (for any fixed $T$) the function $m(T,\cdot)$. It is already shown to be a Weyl--Marchenko function for some $q(\cdot,T)\in\overline{B(-\tau_*^2)}$. Moreover, from (4.10), (4.12)-(4.14) it follows that $q(\cdot,T)$ is reflectionless.

Our first goal is to evaluate the spectrum $\Lambda(q(\cdot,T))$ and (in particular) show that it does not depend on $T$. For this purpose we use the relations between $\Lambda(q(\cdot,T))$ and the set $\Lambda_0(q(\cdot,T))$ of poles of the Weyl--Marchenko function $m(T,\cdot)$ mentioned in the Section 2.
From (4.10), (4.12) we obtain:
$$
\Lambda_0(q(\cdot,T))=\left\{\xi_j(\eta), \ j=\overline{-s,s}, \ \eta\in\Lambda_0\left(Q_T\right)\right\}\cup\{\pm\gamma_\nu\}_{\nu=1}^s. \eqno(4.17)
$$
The total number of the jump points of the function $\sigma(\cdot)$ is $N=(2s+1)n+2s$, where $n=\mbox{card}(\Lambda(Q))$.

Further, since $\vp(\cdot)$ is odd $M(T,-i\eta)=M(T,i\eta)$ implies $m(T,-i\xi_j(\pm\eta))=m(T,i\xi_j(\pm\eta))$, $j=\overline{-s,s}$ and thus for any $\eta\in\Lambda_1(Q_T)$ all corresponding $\left|\xi_j(\eta)\right|, \ j=\overline{-s,s}$ belong to $\Lambda_1(q(\cdot,T))$. The same arguments show that for any $\eta\in\Lambda_2(Q_T)$ all the $\left|\xi_j(\eta)\right|, \ j=\overline{-s,s}$ belong to $\Lambda_2(q(\cdot,T))$. Furthermore, (4.17) shows that all $\gamma_\nu, \ \nu=\overline{1,s}$ belong to $\Lambda_2(q(\cdot,T))$ and (4.2) shows that $m(T,-i\delta_\nu)=m(T,i\delta_\nu)$$=-(g(d_\nu))^{-1}w(T)$ and therefore all $\delta_\nu, \ \nu=\overline{1,s}$ belong to $\Lambda_1(q(\cdot,T))$. Now if we count all the points that are already shown to belong to $\Lambda(q(\cdot,T))$, we obtain $(2s+1)n+s+s=N$. This means that we have found all the elements of this set and we have finally:
$$
\Lambda_1(q(\cdot,T))=\left\{\left|\xi_j(\eta)\right|, \ j=\overline{-s,s}, \ \eta\in\Lambda_1(Q_T) \right\}\cup\{\delta_\nu\}_{\nu=1}^s, \eqno(4.18)$$ $$ \Lambda_2(q(\cdot,T))=\left\{\left|\xi_j(\eta)\right|, \ j=\overline{-s,s}, \ \eta\in\Lambda_2(Q_T)\right\}\cup\{\gamma_\nu\}_{\nu=1}^s. \eqno(4.19)
$$
From (4.18), (4.19) it follows that:
$$
\Lambda(q(\cdot,T))=\left\{\left|\xi_j(\eta)\right|, \ j=\overline{-s,s}, \ \eta\in\Lambda(Q_T)=\Lambda(Q) \right\}\cup\{\delta_\nu\}_{\nu=1}^s \cup\{\gamma_\nu\}_{\nu=1}^s
$$
and consequently the set $\Lambda(q(\cdot,T))$ does not depend upon $T$.

Our next goal is to observe the evolution in $T$ of the normalizing constants $\alpha(\xi,T), \ \xi\in\Lambda:=\Lambda(q(\cdot,T))$. This consideration is based on the relations (2.6), (2.7) applied to both $q(\cdot,T)$ and $Q_T$. There are four different types of eigenvalues that require to be considered separately.

\medskip
{\it Case 1.} $\xi^0=\left|\xi_k(\eta^0)\right|$, $\eta^0\in\Lambda_1(Q_T)$. In this case we use the relation (2.6). From (4.17) and (4.19) we get
$$
\Lambda^1_0(q(\cdot,T))=\left\{\xi_j(\eta), \ j=\overline{-s,s}, \ \eta\in\Lambda^1_0\left(Q_T\right)\right\}
$$
and the relation (2.6) yields:
$$
\alpha(\xi^0,T)=\prod_{\xi\in\Lambda^1_0(q(\cdot,T))}\frac{\xi^0+\xi}{\xi^0-\xi}=
\prod_{\eta\in\Lambda^1_0(Q_T)}\prod_{j=-s}^s\frac{\xi^0+\xi_j(\eta)}{\xi^0-\xi_j(\eta)} =\prod_{\eta\in\Lambda^1_0(Q_T)}\frac{\vp(i\xi^0)+i\eta}{\vp(i\xi^0)-i\eta}.
$$
From this we obtain
$$
\alpha(\xi^0,T)=\prod_{\eta\in\Lambda^1_0(Q_T)}\frac{\eta^0+\eta}{\eta^0-\eta}=A(\eta^0,T),
$$
if $i\eta^0=\vp(i\xi^0)$,
$$
\alpha(\xi^0,T)=\prod_{\eta\in\Lambda^1_0(Q_T)}\frac{-\eta^0+\eta}{-\eta^0-\eta}=A^{-1}(\eta^0,T),
$$
if $i\eta^0=-\vp(i\xi^0)$ and in both cases $A(\eta^0,T)$ denotes the normalizing constant in (2.5) for the potential $Q_T$. Since $A(\eta^0,T)=A(\eta^0,0)\mbox{e}^{-2\eta^0 T}$ we obtain $\alpha(\xi^0,T)=\alpha(\xi^0,0)\mbox{e}^{2i\vp(i\xi^0)T}$.

\medskip
{\it Case 2.} $\xi^0=\delta_k$. Proceeding as above we obtain
$$
\alpha(\delta_k,T)=\prod_{\eta\in\Lambda^1_0(Q_T)}\frac{\vp(i\delta_k)+i\eta}{\vp(i\delta_k)-i\eta}.
$$
Since $\vp(i\delta_k)=0$ this yields: $\alpha(\delta_k,T)=\alpha(\delta_k,0)$ that can be written in the same form as in case 1: $\alpha(\delta_k,T)=\alpha(\delta_k,0)\mbox{e}^{2i\vp(i\delta_k)T}$.

\medskip
{\it Case 3.} $\xi^0=\left|\xi_k(\eta^0)\right|$, $\eta^0\in\Lambda_2(Q_T)$. In this case we use the relation (2.7) that yields:
$$
\alpha(\xi^0,T)=-\frac{d\sigma(-\xi^0,T)}{d\sigma(\xi^0,T)}\prod_{\xi\in\Lambda^1_0(q(\cdot,T))}\frac{\xi^0+\xi}{\xi^0-\xi}= -\frac{d\sigma(-\xi^0,T)}{d\sigma(\xi^0,T)}\prod_{\eta\in\Lambda^1_0(Q_T)}\prod_{j=-s}^s\frac{\xi^0+\xi_j(\eta)}{\xi^0-\xi_j(\eta)} = $$ $$-\frac{d\sigma(-\xi^0,T)}{d\sigma(\xi^0,T)}\prod_{\eta\in\Lambda^1_0(Q_T)}\frac{\vp(i\xi^0)+i\eta}{\vp(i\xi^0)-i\eta}.
$$
Using (4.10) we obtain:
$$
\alpha(\xi^0,T)=-\frac{d\theta(-\eta^0,T)}{d\theta(\eta^0,T)}\prod_{\eta\in\Lambda^1_0(Q_T)}\frac{\eta^0+\eta}{\eta^0-\eta}=A(\eta^0,T),
$$
if $i\eta^0=\vp(i\xi^0)$,
$$
\alpha(\xi^0,T)=-\frac{d\theta(\eta^0,T)}{d\theta(-\eta^0,T)}\prod_{\eta\in\Lambda^1_0(Q_T)}\frac{-\eta^0+\eta}{-\eta^0-\eta}=A^{-1}(\eta^0,T),
$$
if $i\eta^0=-\vp(i\xi^0)$. In both cases we get $\alpha(\xi^0,T)=\alpha(\xi^0,0)\mbox{e}^{2i\vp(i\xi^0)T}$.

\medskip
{\it Case 4.} Consider the points $\gamma_k, k=\overline{1,s}$. Here (2.7) yields:
$$
\alpha(\gamma_\nu,T)=-\frac{d\sigma(-\gamma_\nu,T)}{d\sigma(\gamma_\nu,T)}\prod_{\xi\in\Lambda^1_0(q(\cdot,T))}\frac{\gamma_\nu+\xi}{\gamma_\nu-\xi}. \eqno(4.20)
$$
First, from (4.12) we obtain:
$$
-\frac{d\sigma(-\gamma_\nu,T)}{d\sigma(\gamma_\nu,T)}=\frac{w(T)-M(T,(-1)^\nu i\kappa^*)}{w(T)-M(T,(-1)^{\nu-1} i\kappa^*)}. \eqno(4.21)
$$
Since $w(t)$ can be represented as
$$
w(t)=\frac{\dot z(t)}{z(t)},
$$
where $z(t)$ is a solution of the Cauchy problem
$$
-\ddot z+Q(t)z=\mu^* z, \ z(0)=1, \ \dot z(0)=w_0
$$
we can rewrite (4.21) into the following form:
$$
-\frac{d\sigma(-\gamma_\nu,T)}{d\sigma(\gamma_\nu,T)}=\mbox{const}\cdot \frac{\psi(T,(-1)^{\nu-1}i\kappa^*)}{\psi(T,(-1)^\nu i\kappa^*)}, \eqno(4.22)
$$
where $\psi(t,\rho)$ is the Weyl--Marchenko solution for $Q$.

On the other hand, the Jost solution $e^+(t,\rho)$ for $Q$ admits the representation {\cite{march}}:
$$
e^+(t,\rho)=\mbox{e}^{i\rho t}\frac{\prod\limits_{\eta\in\Lambda_0(Q_t)}(\rho-i\eta)}{\prod\limits_{\eta\in\Lambda(Q)}(\rho+i\eta)}, \eqno(4.23)
$$
that yields, in particular, for any $\tau:\ \pm\tau\notin\Lambda(Q)\cup\Lambda_0(Q_T)$:
$$
\prod_{\eta\in\Lambda_0(Q_T)}\frac{\tau+\eta}{\tau-\eta}=C(\tau)\cdot\mbox{e}^{2i(i\tau) T}\frac{e^+(T,-i\tau)}{e^+(T,i\tau)}. \eqno(4.24)
$$
Gathering together the relations (4.20)-(4.24) we obtain:
$$
\alpha(\gamma_\nu,T)=B_\nu \cdot\frac{\psi(T,(-1)^{\nu-1}i\kappa^*)}{\psi(T,(-1)^\nu i\kappa^*)}\cdot\mbox{e}^{2i(-1)^{\nu-1}i\kappa^* T}\frac{e^+(T,(-1)^{\nu}i\kappa^*)}{e^+(T,(-1)^{\nu-1}i\kappa^*)}
$$
with some constant $B_\nu$.
Since Jost $e^+(t,\rho)$ and Weyl--Marchenko $\psi(t,\rho)$ solutions for reflectionless potentials are proportional, finally this yields $\alpha(\gamma_\nu,T)=\alpha(\gamma_\nu,0)\mbox{e}^{2i\vp(i\gamma_\nu)T}$.

Thus we have $\alpha(\xi,T)=\alpha(\xi,0)\mbox{e}^{2i\vp(i\xi)T}$ for all $\xi\in\Lambda(q(\cdot,T))=\Lambda(q(\cdot,0))$. Repeating the similar arguments as in proof of the Theorem 35.19 {\cite{BeDT}} we can conclude that $q$ solves the equation (3.1).
$\hfil\Box$

\medskip
In order to complete the proof of Theorem 4.1 we take an arbitrary $Q\in\overline{B(\mu_*)}$ and define $m(t,\cdot)$ via (4.2). It follows from Lemma 4.1 that for any fixed $t$ $m(t,\cdot)$ is a Weyl--Marchenko function for some $q(\cdot,t)\in\overline{B(-\tau_*^2)}$ and $q$ satisfies the boundary conditions (3.2). Let us show that it also solves the equation (3.1).

Consider the sequence $Q_N\in B(\mu_*)$ convergent to $Q$ in the topology of $\overline{B(\mu_*)}$ (i.e., in the topology of uniform convergence of the functions and all their derivatives on any compact set) such that the corresponding Weyl--Marchenko functions $M_N(0,\rho)$ converge to the Weyl--Marchenko function $M(0,\rho)$. Such sequence exists by virtue of the Proposition 2.1, moreover, since the Weyl--Marchenko solutions $\psi_N(t,\rho)$ and $\psi(t,\rho)$ with $\rho\in\mathbf{C}\setminus[-i\kappa_*,i\kappa_*]$ can not vanish for any $t$ (this was mentioned at the beginning of this proof), Remark 2.1 guarantees that $M_N(t,\rho)=(\psi_N(t,\rho))^{-1}\dot\psi_N(t,\rho)$ converge to $M(t,\rho)=(\psi(t,\rho))^{-1}\dot\psi(t,\rho)$ for any fixed $t$. Let us define
$$
m_N(t,\rho):=\frac{M_N(t,\vp(\rho))-w_N(t)}{g\left(\rho^2\right)},
$$
where $w_N$ is a solution of the Cauchy problem:
$$
\dot{w}_N+w_N^2=Q_N(t)-\mu^*, \quad w_N(0)=w_0.
$$
Since for sufficiently large $N$ we have $w_0\in (M_N(0,i\kappa^*),M_N(0,-i\kappa^*))$ all $w_N(t)$ with sufficiently large $N$ are finite for all $t$ and $w_N(t)\to w(t)$ as $N\to\infty$. Thus we have
$$
m(t,\rho)=\lim\limits_{N\to\infty}m_N(t,\rho), \ \rho\in\mathbf{C}\setminus[-i\tau_*, i\tau_*]. \eqno(4.25)
$$
On the other hand, by virtue of Lemmas 4.1, 4.2 $m_N(t,\cdot)$ are the Weyl--Marchenko functions for $q_N(\cdot,t)\in B(-\tau_*^2)$, where $q_N$ satisfies the equation (3.1).

Let $\mathcal{B}(-\tau_*^2)$ be the set of all the solutions of (3.1) that belong to $B(-\tau_*^2)$ for each fixed $t$ considered with the topology of the uniform convergence of the functions with all their derivatives  on any compact set of the $(x,t)$ - plane. As in proof of {\cite{march}}, Theorem 2.3 $\mathcal{B}(-\tau_*^2)$ can be shown to be a precompact set. So there exists $q^*$ such that some subsequence $q_{N_n}(x,t)$ converges to $q^*(x,t)$ as $n\to\infty$ together with all the derivatives uniformly on any compact set of the $(x,t)$ - plane. Clear that $q^*$ satisfies the equation (3.1). At the same time it is clear that for any fixed $t$ $q_{N_n}(\cdot,t)\to q^*(\cdot,t)$ in the topology of $\overline{B(-\tau_*^2)}$. This means that  $q^*(\cdot,t)\in\overline{B(-\tau_*^2)}$ for each fixed $t$. In view of the Proposition 2.1 for each fixed $t$ there exists the subsequence $n(k)$ such that the corresponding Weyl--Marchenko function $m^*(t,\rho)=\lim\limits_{k\to\infty}m_{N_{n(k)}}(t,\rho)$. Together with (4.25) this yields $m^*(t,\rho)\equiv m(t,\rho)$, consequently, $q^*=q$ and the Theorem 4.1 is proved.
$\hfil\Box$

\smallskip
{\bf Remark 4.1} It follows from Lemmas 4.1, 4.2 that the procedure presented in the Theorem 4.1 allows to construct, in particular, the soliton solutions for the problem (3.1), (3.2): it is sufficient to choose $Q$ from the proper class of reflectionless potentials. In an analogous way the finite-gap solutions for the problem can be constructed via the same procedure. For this purpose one should choose $Q$ as a finite-gap potential with all the gaps lying on the interval $(\mu_*,0)$ and set $w_0=M(0, -i\kappa^*)$ or $w_0=M(0, i\kappa^*)$ (one can easily show that the assertion of the Theorem 4.1 remains true in this case).

\smallskip
{\bf Remark 4.2} In our considerations we treated the function $Q$ as a free parameter but it can also be described in terms of boundary values of the solution $q(x,t)$. Namely, as it follows from the proof of the Lemma 4.1, $w(t)$ (which is a solution of the Cauchy problem (4.1)) can be written as $w(t)=(-1)^{s-1}4^s b_{2s}\left(q(\cdot,t)\right)$.

\medskip
{\bf Acknowledgment.} This work was supported by the Russian Ministry of
Education and Science (Grant 1.1436.2014K).

%\section*{References}

\end{document}